\documentclass[aps,showpacs,pra,twocolumn]{revtex4}
\usepackage{mathrsfs} 
\usepackage{amssymb}

\usepackage{graphicx}
\usepackage{dcolumn}
\usepackage{bm}

\begin{document}

\title{Non-classical correlation of cascaded photon pairs emitted from quantum dot}
\author{Chuan-Feng Li$\footnote{
email: cfli@ustc.edu.cn}$, Yang Zou, Jin-Shi Xu, Rong-Chun Ge and Guang-Can Guo} \affiliation{Key Laboratory of Quantum Information, University of Science and
Technology of China, CAS, Hefei, 230026, People's Republic of China}
\date{\today }


\begin{abstract}
We studied the quantum correlation of the photon pairs generated by biexciton cascade decays of self-assembled quantum dots, and determined the correlation
sudden-change temperature, which is shown to be independent of the background noise, far lower than the entanglement sudden-death temperature, and therefore, easier
to be observed in experiments. The relationship between the fine structure splitting and the sudden-change temperature is also provided.
\end{abstract}

\maketitle

\section{Introduction}

It is impossible to overestimate the role, which entangled photon pairs have played and continue to play in the field of quantum communication and quantum
information \cite{Gisin2002rmp,Gisin2007nat}. Polarization entangled photon pairs are routinely produced by nonlinear optical effects, predominantly by parametric
down-conversion process \cite{Bouwmeester1997nat}, of which entanglement dynamics has been investigated, and many extraordinary results have been obtained, such as
entanglement sudden death (ESD) \cite{Yu2004prl,Yu2006prl,Xu2009prl}, and sudden birth \cite{Xu2009prl,Vedral2009jp}, etc.

The proposal that the biexciton-radiative cascade process in a single quantum dot (QD) provides a source of polarization entangled photon pairs was first made by
Benson {\it et al.} \cite{Benson2000prl}, and has been realized by Stevenson {\it et al.} and Akopian {\it et al.} \cite{Akopian2006prl,Stevenson2006nature}.
Several tests have also been performed on the state of the photon pair to study the entanglement behavior \cite{Young2006njp,Young2009prl}.

However, recent works show that, even without entanglement, there are quantum tasks \cite{Lanyon2008prl,Meyer2000prl,Datta2008prl} superior to their classical
counterparts. This is due to their nonzero quantum correlations, introduced by Henderson and Vedral \cite{Henderson2001jp}, which describes all the non-classical
correlations in a state, while entanglement is only a special part of the correlations. Recently, some special results have been obtained for quantum correlation,
such as sudden change \cite{Maziero2009pra,Xu2010nat,Mazzola2010prl} in the evolution, that is the derivation of quantum correlation is not continuous at some
points as it evolves with time.

In this work, we use the model that we presented in Ref.\,\cite{Zou2010pra} to investigate quantum correlation dynamics of the photon pairs generated from a QD.
When we consider the influence of the temperature of the QD system, a special temperature has been found, at which the behavior of quantum correlation suddenly
changes. Our study also indicates that this temperature is independent of the background noise and lower than entanglement sudden-death temperature. Therefore, the
sudden-change behavior of correlation may be easier to be observed in experiments.

The paper is organized as following. In Sec.\,\ref{Model and Method}, we briefly introduce the model of QD and the method that we used to
calculate the quantum correlation of the photon pair. The result is described in Sec.\,\ref{result}. We present our conclusion in
Sec.\,\ref{conclusion}.

\section{Model and Method}\label{Model and Method}

\begin{figure}[b]
\centering
\includegraphics[width= 3in]{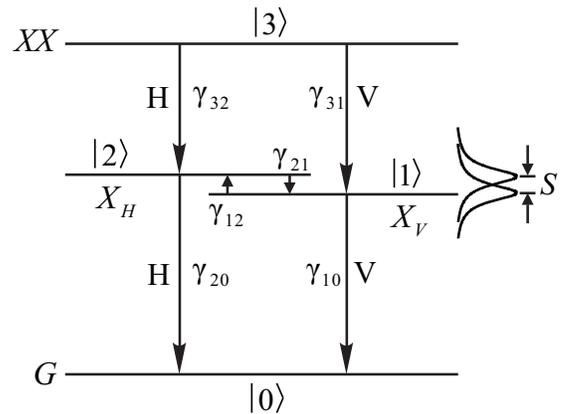}
\caption{Energy level schematic of the biexciton cascade process, the ground state $G$ ($|0\rangle$), the two linear polarized exciton state $X_{H}$ ($|2\rangle$)
and $X_{V}$ ($|1\rangle$) and the biexciton state $XX$ ($|3\rangle$). The spontaneous emission process is marked by $\gamma_{ij}$ ($\gamma_{12}$ and $\gamma_{21}$
are phonon assisted transition rates).}\label{fourlevel}
\end{figure}

\begin{figure*}[t]
\centering
\includegraphics[width= 6in]{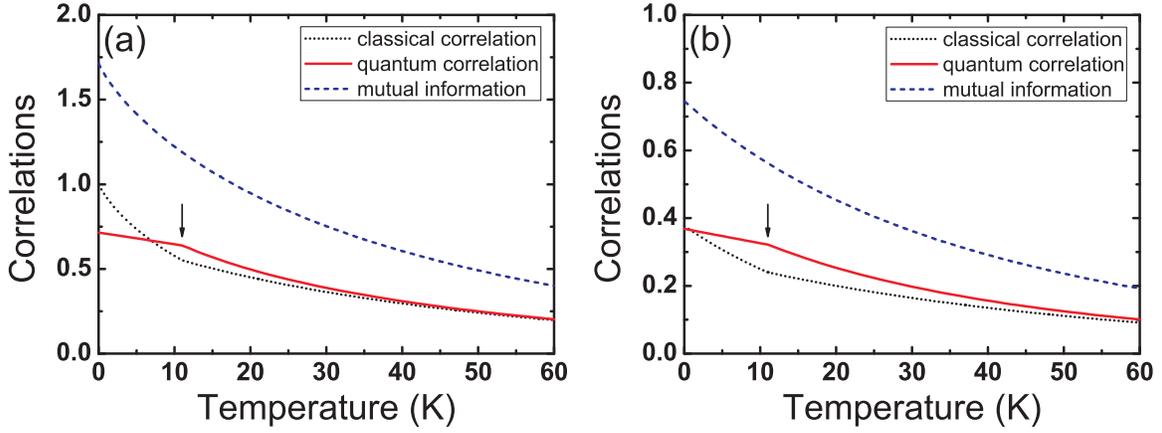}
\caption{(color online). Values of the classical correlation (dotted line), quantum correlation (solid line), and the mutual information (dashed
line) as a function of temperature with fixed evolution time $\tau_g=0.5$\,ns, and gate width $w_g=0.1$\,ns. (a) without background noise, $g=0$
(b) with background noise, $g=0.45$.}\label{CT}
\end{figure*}

The model of the biexciton cascade process in a QD is presented in Figure\,\ref{fourlevel}. A single QD is initially excited to the biexciton state, involving two
electrons and two holes by a short-pulsed laser, and subsequently evolves freely. A biexcton photon $H_{XX}$ or $V_{XX}$ is emitted as the dot decays to an exciton
($X$) state by recombining one electron and one hole. The polarization of the biexciton photon is either horizontal ($H$) or vertical ($V$), in accord with the
decay into the exciton state $X_H$ or $X_V$, respectively. After some time delay $\tau$, the other electron and hole recombine to emit an exciton photon $H_X$ or
$V_X$ with the same polarization as that of the earlier biexciton photon. Therefore, this process generates entangled two-photon state \cite{Stevenson2008prl}
\begin{equation}
|\Psi\rangle = \frac{1}{\sqrt{2}}(|H_{XX}H_X\rangle + e^{iS\tau}|V_{XX}V_X\rangle)\label{idealstate}
\end{equation}
with $S$ the energy splitting between the exciton states, as shown in Fig.\,\ref{fourlevel}, which has conventionally been called the ``fine
structure splitting" (FSS), and can be tuned by a lot of methods in experiment \cite{Young2005prb,Seidl2006apl,Stevenson2006prb,Marcet2010apl}.

\begin{figure*}[tbph]
\centering
\includegraphics[width = 6in]{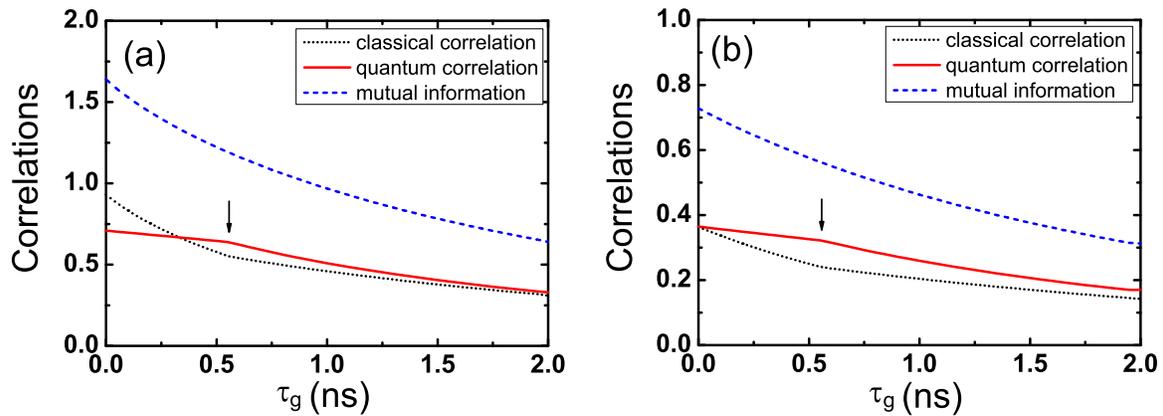}
\caption{(color online). Values of the classical correlation (dotted line), quantum correlation (solid line), and the mutual information (dashed
line) as a function of emission delay $\tau_{g}$ at the temperature of $10$K. (a) without noise (b) with noise.}\label{Ctao}
\end{figure*}

Due to this FSS being about tens of $\mu$eV, it is necessary to consider the phonon assisted transition between the two exciton states ($X_H$ and $X_V$) in the
process. In this system, the phonon absorption rate is $\gamma_{12}=\kappa N_{B}$ and the emission rate is $\gamma_{21}=\kappa(N_{B}+1)$, where $\kappa$ is the
phonon-QD interaction rate, which is approximately proportional to the cube of the energy splitting $S$ \cite{Shen2007prb}, and $N_{B}$ is the Bose distribution
function of a phonon with energy $S$, $N_{B}=[\exp(S/k_{B}T)-1]^{-1}$. In experiment, in order to overcome the influence of FSS, a time gate can be used to select
the exciton photon with only a short emission delay $\tau_{g}\leq\tau\leq(\tau_{g}+w_{g})$ relative to the biexciton photon, which has been carried out by Stevenson
{\it et al.} \cite{Stevenson2008prl}.

In our calculation, we assume the total polarization density matrix $\hat\rho_{\text{pol}}$ of the cascaded emission photon includes three parts: the one with
non-classical correlations $\hat\rho_1$ whose portion $\eta$ is determined by the spectrum overlap of exciton and biexciton photons, the distinguished part
$\hat\rho_2$ and the background noise $\hat\rho_3$,
\begin{eqnarray}
\hat{\rho}_{\text{pol}} = \frac{1}{1+g}[\eta\hat{\rho}_{\text{1}} + (1-\eta)\hat{\rho}_{\text{2}} + g\hat{\rho}_{\text{3}} \label{rhot}].
\end{eqnarray}
The elements of $\hat\rho_1$ can be derived by master equation method and quantum regression theorem which is presented in Ref.\,\cite{Zou2010pra}. The second term
$\hat\rho_2$ has the same diagonal elements with $\hat\rho_1$, but its nondiagonal elements are all zero. The noise term $\hat\rho_3$ is set as an identity matrix.

To obtain quantum correlation, classical correlation must be acquired first, which is defined as \cite{Maziero2009pra,Vedral2003prl}
\begin{eqnarray}\label{classical}
\mathcal C(\hat{\rho}_{\text{pol}})\equiv \max_{\{\Pi_j\}}[S(\hat{\rho}_{_{XX}})-S_{\{\Pi_j\}}(\hat\rho_{_{XX|X}})],
\end{eqnarray}
where the maximum is taken over the set of projective measurements $\{\Pi_j\}$ on the subsystem of exciton photon,
$S_{\{\Pi_j\}}(\hat\rho_{_{XX|X}})=\sum_j q_j S(\hat\rho_{_{XX}}^j)$ is the conditional entropy of the subsystem of biexciton photon, given the
knowledge (measure) of the state of the exciton photon, $\hat\rho_{_{XX}}^j=\text{Tr}_{_X}(\Pi_j\hat\rho_{\text{pol}}\Pi_j)/q_j$,
$q_j=\text{Tr}(\hat\rho_{\text{pol}}\Pi_j)$, $\hat\rho_{_{XX(X)}}=\text{Tr}_{_{XX(X)}}(\hat\rho_{\text{pol}})$ is the reduced density matrix of
the biexciton (exciton) photon, and $S(\hat\rho_{_{XX(X)}})\equiv-\text{Tr}(\hat\rho_{_{XX(X)}}\log_2\hat\rho_{_{XX(X)}})$ is the Von Neumann
entropy of $\hat\rho_{_{XX(X)}}$. Then quantum correlation is obtained by subtracting $\mathcal C$ from the quantum mutual information which
denotes the total correlation, that is
\begin{eqnarray}\label{quantum}
\mathcal Q(\hat\rho_{\text{pol}})=\mathcal I(\hat\rho_{\text{pol}})-\mathcal C(\hat\rho_{\text{pol}}),
\end{eqnarray}
where $\mathcal I(\hat\rho_{\text{pol}})=S(\hat\rho_{_{XX}})+S(\hat\rho_{_{X}})-S(\hat\rho_{\text{pol}})$.

In all the calculations, the temporal window width $w_g$ is fixed at $0.1$\,ns, short enough to select entangled photon pairs. Because this short temporal window is
equivalent to about $41$\,$\mu$eV in the energy domain according to Fourier transform, which is far larger than the FSS that we set, thus it can not resolve the
which-path information within the cascade emission process.

\section{Results}\label{result}

The correlations dynamic with respect to the temperature is shown in Fig.\,\ref{CT}. We set FSS as $2.5$\,$\mu$eV and fix the emission delay $\tau_g$ at $0.5$\,ns,
which is shorter than the exciton lifetime ($\sim$ 0.8\,ns) and can be recognized as the evolution time of the system. We compare two conditions: (a) without
background noise $(g=0)$ and (b) with a large background noise $(g=0.45)$. The factor $g$ discussed in Ref.\,\cite{Zou2010pra} represents the portion of the noise.
In both of the conditions, the correlations of the photon pair reduce monotonically with temperature, due to the faster phonon assisted transition rate $\gamma_1$
and $\gamma_2$ at raised temperature resulting in the larger decoherence.

\begin{figure}[b]
\centering
\includegraphics[width= 3.1in]{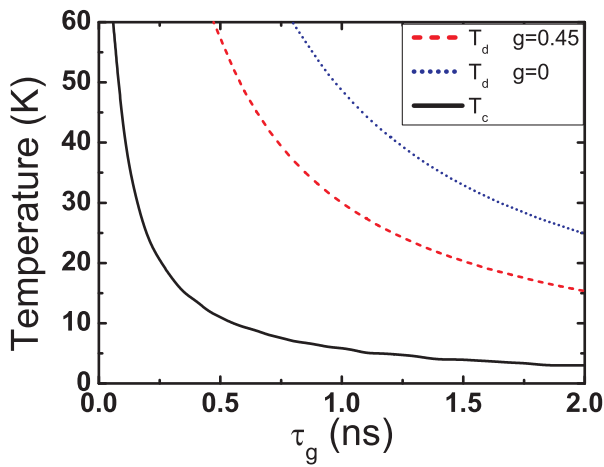}
\caption{(color online). Critical temperatures as a function of emission delay. $T_{c}$ and $T_{d}$ represent the sudden-change temperature of correlation and
sudden-death temperature of entanglement, respectively. The other parameters are FSS $S=2.5$\,$\mu$eV, and temporal window $w_g=0.1$\,ns. }\label{Tt}
\end{figure}

An interesting phenomenon is that the quantum correlation decreases slowly at low temperature, but the decreasing rate changes to be fast suddenly when the system
reaches to a critical temperature, $T_c$, indicated by the arrow in the figures, which we call the sudden-change temperature. At $T_c$, the first order derivative
of the function of the quantum correlation behavior is not continuous. Moreover, $T_c$ is almost the same for both of the two conditions, indicating that the
evolution behavior of quantum correlation is independent of the background noise. This can be understood from the definition of quantum correlation in
Eqs.\,(\ref{classical}), and (\ref{quantum}). The noise part of the matrix is an identity matrix, which only reduces the correlation, but does not influence its
derivation. Here we also notice that the quantum correlation may be greater than the classical one at some temperatures, for example, at the critical temperature.

\begin{figure}[b]
\centering
\includegraphics[width= 3in]{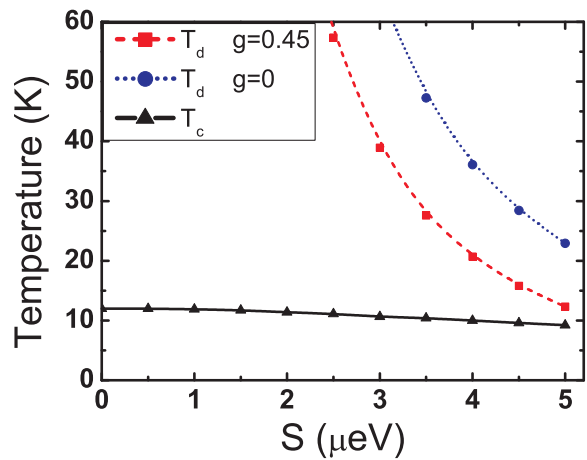}
\caption{(color online). Correlation sudden change temperature ($T_{c}$) and entanglement sudden death temperature ($T_{d}$) as a function of FSS. Emission delay
$\tau_g=0.5$\,ns, and temporal window $w_g=0.1$\,ns.}\label{TS}
\end{figure}

In Fig.\,\ref{Ctao}, the behavior of correlations with increased emission delay $\tau_{g}$ is similar to that in Fig.\,\ref{CT}. No matter whether there is
background noise, the phenomenon of quantum correlation sudden-change still takes place at the same point of emission delay.

It is worth mentioning that this peculiar sudden-change behavior is different from ESD. Indeed, we have shown this system also undergoes ESD \cite{Zou2010pra}, but
at a different temperature $T_d$.

To see more about the correlation sudden-change temperature $T_c$ and ESD temperature $T_d$, we plot them as a function of the emission delay $\tau_g$, as shown in
Fig.\,\ref{Tt}, where $T_d$ is found much higher. For $\tau_g$ near to zero, both of the critical temperatures are higher than $60$\,K, and they decrease rapidly.
At the beginning, $T_c$ decreases faster, but when the emission delay is longer than $0.5$\,ns, the decreasing rate apparently becomes slow. Moreover, as we have
mentioned above, $T_c$ is independent of the background noise, but $T_d$ is lower for larger noise.

In Fig.\,\ref{TS}, we show the critical temperatures depending on FSS. Obviously, the entanglement sudden-death temperature decreases fast from a large value,
however, the correlation sudden-change temperature is always around $10$\,K, and decreases slightly. All the phenomenon indicate that quantum correlation
sudden-change temperature can be hardly influenced by the non-quantum factors, but is determined by the phonon assisted process.

Comparing the phenomenon of correlation sudden change and ESD in this system, we find the former may be easier to be observed in experiment, because it is weakly
dependent on the noise and FSS, and the critical temperature is much lower resulting in better signal to noise ratio to reconstruct the polarization matrix of the
photon pair by means of quantum state tomography \cite{James2001pra}. However, to observe ESD, a large FSS is needed, which makes the degree of entanglement
extremely low and request much higher accuracy in experiment.

\section{Conclusion}\label{conclusion}

In conclusion, we have calculated the quantum correlation of the photon pair emitted from quantum dots, and considered how the temperature, evolution time and FSS
affect the quantum correlation of the photon pair. We found the phenomenon that quantum correlation behavior suddenly changed at a critical temperature. This
critical temperature is independent of the background noise in the system and is weakly dependent on the FSS of a QD. Moreover, the temperature is low enough for QD
to emit cascaded photon pair with more correlation and entanglement, which makes sudden-change of correlation easier to be observed in experiments.

This work was supported by National Basic Research Program, and National Natural Science Foundation of China (Grant Nos. 60921091, 10874162 and 10734060).

\end{document}